\documentclass[preprint,nofootinbib]{revtex4}

\usepackage{graphicx}

\begin{document}

\title{Stability of the Einstein static universe in presence of vacuum energy}

\author{Saulo Carneiro$^{1,2}$\footnote{ICTP Associate Member. E-mail address: saulo.carneiro@pq.cnpq.br.} and
Reza Tavakol$^{1}$\footnote{E-mail address: r.tavakol@qmul.ac.uk.}}

\affiliation{$^1$Astronomy Unit, School of Mathematical Sciences,
Queen Mary University of London, Mile End Road, London E1 4NS,
UK\\$^2$Instituto de F\'{\i}sica, Universidade Federal da Bahia,
Salvador, BA, 40210-340, Brazil}

\begin{abstract}
The Einstein static universe has played a central role in a number of
emergent scenarios recently put forward to deal with the
singular origin of the standard cosmological model.
Here we study the existence and stability of the
Einstein static solution in presence of vacuum energy corresponding
to conformally-invariant fields.
We show that the presence of vacuum energy
stabilizes this solution by changing it
to a centre equilibrium point, which is cyclically stable.
This allows non-singular emergent cosmological models
to be constructed in which initially the universe oscillates
indefinitely about an initial Einstein static solution
and is thus past eternal.
\end{abstract}

\maketitle

\section{Introduction}
Recent accumulation of high resolution observations is compatible
with the so called standard model of cosmology which has a number of
intriguing features. In addition to the early and late accelerating
phases, which are difficult to account for within the classical
relativistic framework with non-exotic matter sources, this model also
possesses an initial singular state at which the laws of physics
break down. To deal with this latter shortcoming, a number of
attempts have recently been made to construct models which are
non-singular and/or past eternal\footnote{The history of attempts at
constructing non-singular/oscillatory universes which are past
eternal goes much further back to at least the work of Tolman
\cite{Tolman} (see also \cite{novello} and references therein for a
review of non-singular models).}. These fall into a number of
groups, including emergent scenarios \cite{EM,GMT,TL} (see also
\cite{Carneiro,CT}) and cyclic/ekpyriotic models \cite{cyclic}.
There are also a number of other cosmological models
which are not necessarily  recurrent, but nevertheless are
non-singular in the past. These include the pioneering
non-singular model by Bojowald based on modifications due to Loop Quantum
Gravity \cite{Bojowald}, the model
based on polymer matter \cite{Hossain} and others \cite{Sahni}.

The so called emergent scenarios, which are
non-singular and past eternal, come
in two varieties: those that employ
classical general relativity \cite{EM,GMT},
and those that incorporate
quantum effects which are expected to be
present at early phases of the universe \cite{TL}.
An important ingredient in the construction of
these models is the Einstein static solution,
which in the classical general relativistic setting
has long been known to be unstable\footnote{This
solution has, however, been shown to be stable with respect to
inhomogeneous perturbations \cite{StableEinsteinStatic}).}.
This instability makes the construction of emergent models
within the framework of classical relativity difficult.
Interestingly though, quantum effects which
are known to be operative at early phases of
the universe have recently been shown to be able to
stabilise the Einstein static solution by
changing it from a hyperbolic equilibrium point (which is
unstable) to a centre equilibrium point, which is cyclically
stable. This was first shown to be the case in
presence of quantum modifications due to Loop Quantum
Gravity effects in \cite{TL} (see also \cite{Vereshchagin,mntl}
and other related works in this connection \cite{EU,Parisi}).
In most cases, where the
singularity is removed, the overall effect is to
change the effective evolution equations in such a way
which allow the singularity theorems to be circumvented.
For example, in the context of Loop Quantum Cosmology
it has been shown (see \cite{TL} for details) that,
by writing the effective evolution equations
in terms of an effective fluid, the equation of state in the
semi-classical regime can change, and even become phantom
(with equation of state parameter $<-1$). This results in the
violation of the null energy condition which removes
a crucial barrier to singularity avoidance.

Here we consider an ingredient not studied before, namely
the contribution of vacuum quantum fluctuations to the energy content of classical Einstein equations,
and study their effect on the
existence and stability of the Einstein static solution.
In this connection we recall an
earlier result \cite{Ford},
according to which the vacuum energy of conformal fields
in a spatially closed static universe
has a density proportional to $a^{-4}$,
where $a$ is the scale factor.
This result has been confirmed more recently by other studies
which show that the Casimir energy in such backgrounds is
proportional to $a^{-1}$ \cite{Elizalde}.
We study the effects of presence of
this vacuum energy on the dynamics of the universe and
show that it has the consequence of stabilising the
Einstein static solution by changing it
into a centre equilibrium point.

The outline of the paper is as follows.
In Section \ref{general} we perform a general
analysis of the equilibrium of Einstein solution in the presence of matter and the vacuum energy. In Section \ref{numerical} we consider a
numerical example, in which the energy content is assumed to consist of relativistic
matter plus a vacuum term with negative pressure.
In Section \ref{scalarfield} we study the existence of Einstein static solution
in presence of a single scalar field, with the vacuum contribution
represented by a self-interaction potential.
We then briefly discuss possible alternatives for the exit from the initial Einstein phase
to an inflationary one. We conclude with a discussion in Section \ref{conclusions}.

\section{Study of the Einstein universe in the general setting}

\label{general}

In this section we study the effects
of vacuum energy due to conformal fields
on the dynamics of the universe, and
in particular the way this affects the
existence and the stability of the Einstein static solution.
As was mentioned above the
vacuum energy of a conformal scalar field
in a spatially closed static universe of radius $a$
has been shown to have a density given by
$\rho_\Lambda = C/a^4$, where $C$ is a positive constant \cite{Ford}.
Initially it was also proposed that the corresponding
pressure should have the form  $p_{\Lambda} = \rho_{\Lambda}/3$,
to ensure the vacuum energy-momentum tensor is traceless
in order to respect the conformal symmetry.
However, with the discovery of trace anomaly such a requirement is
not necessary. In the de Sitter space-time the
equation-of-state parameter of the vacuum is $\omega_{\Lambda} = -1$,
due to the symmetry of the background.
This result, however, cannot be assumed to hold in
other more general space-times.
Here, therefore,
we shall proceed by first considering the general case,
$p_\Lambda = \omega_{\Lambda} \rho_\Lambda$,
where
$\omega_{\Lambda}$ is allowed to take arbitrary values.
We shall then
find the conditions for the stability of the
Einstein static universe in terms of $\omega_{\Lambda}$.

Starting with a closed
isotropic and homogeneous Friedmann-Lema\^{\i}tre-Robertson-Walker
model sourced by a general fluid with total density $\rho$ and
total pressure $p$, the evolution equations are given by
\begin{eqnarray}
3H^2 = \rho - \frac{3}{a^2}, \label{Friedmann}\\
\dot{\rho}+3H(\rho+p)=0, \label{continuity}
\end{eqnarray}
where $H$ is the Hubble parameter. We shall assume the
fluid to consist of a combination of
the above vacuum energy plus matter with
$p_m = \omega_m \rho_m$,
where $\rho_m$, $p_m$ and $\omega_m$ are the corresponding
density, pressure and equation-of-state parameter.
Substituting these in the above
evolution equations and letting $\rho_{\Lambda} = C/a^4$,
the Raychadhuri equation can be written as
\begin{equation}\label{secondorder}
\ddot{a} = - \frac{\dot{a}^2+1}{2a}\left(1+3\omega_m\right)+\frac{C}{2a^3}\left(\omega_m-\omega_{\Lambda}\right).
\end{equation}

The Einstein static solution is given by
$\ddot a = 0 = \dot a$.
To begin with we obtain the conditions for the existence of
this solution. The scale factor in this case
is given by
\begin{equation}\label{FP}
a_{ES}^2 = \frac{C\left(\omega_m-\omega_{\Lambda}\right)}{3\omega_m+1}.
\end{equation}
The existence condition reduces to the
reality condition for $a_{ES}$, which for a positive $C$ takes the forms
\begin{equation}\label{condition1}
\omega_m > - 1/3\;\;\; \mbox{and}\;\;\;
\omega_{\Lambda} < \omega_m,
\end{equation}
or
\begin{equation}\label{condition2}
\omega_m < - 1/3\;\;\; \mbox{and}\;\;\;
\omega_{\Lambda} > \omega_m.
\end{equation}
Therefore, in the case of ordinary matter ($\omega_m\geq0$) plus a
positive vacuum energy with negative pressure
(i.e. with $\omega_{\Lambda}<0$), the Einstein static solution always
exists\footnote{It is interesting to note that conditions
(\ref{condition1})-(\ref{condition2}) exclude the case
$\omega_{\Lambda} = 1/3$, originally proposed in \cite{Ford}.}.

To study the stability of this solution, it is
helpful to cast the equation (\ref{secondorder})
as a $2$-dimensional dynamical system by introducing
the phase-space variables $x_1=a$ and $x_2=\dot{a}$,
\begin{eqnarray}\label{system1}
\dot{x}_1 &=& x_2,\\ \label{system2}
\dot{x}_2 &=& - \frac{x_2^2+1}{2x_1}\left(1+3\omega_m\right)+\frac{C}{2x_1^3}\left(\omega_m-\omega_{\Lambda}\right).
\end{eqnarray}
In these variables the Einstein static solution corresponds to
the fixed point ($x_1= a_{ES}, ~ x_2=0$).
The stability of this equilibrium point is readily found by
looking at the eigenvalues, $\lambda$,
of the Jacobian matrix $J_{ij} = \partial \dot{x}_i/\partial x_j$
evaluated at this point, which are
found to be
\begin{equation}
\lambda^2 = - \frac{C\left(\omega_m-\omega_{\Lambda}\right)}{a_{ES}^4}.
\end{equation}

The stability depends on the
sign of $\lambda^2$. For $\lambda^2 >0$,
the Einstein static solution is a hyperbolic
fixed point and hence unstable,
in the sense that trajectories starting in the neighbourhood
of such a point exponentially diverge from it
(this is the same as the classical
relativistic case). For $\lambda^2 <0$, on the other hand,
the Einstein static solution
becomes a centre equilibrium point, which is circularly
stable, in the sense that small departures from the fixed point
results in oscillations about that point rather than exponential
deviation from it. In this case the universe
stays (oscillates) in the neighbourhood of the Einstein static
solution indefinitely.
Thus the condition for stability is given by $\lambda^2 <0$.
For $C>0$, this implies that $\omega_{\Lambda}<\omega_m$.
Comparing this inequality with the conditions for existence
of the Einstein static solution, (\ref{condition1})-(\ref{condition2}),
we conclude that, given  $\omega_{\Lambda}<\omega_m$,
the Einstein universe is stable for $\omega_m > -1/3$.
In particular, it is stable in presence of
ordinary matter ($\omega_m \ge 0$) plus a positive vacuum energy with negative
pressure\footnote{In the case of a negative vacuum energy, that is $C<0$,
the conditions for existence and stability of the Einstein solution are
given by $\omega_{\Lambda}>\omega_m>-1/3$, which are satisfied by any
ordinary matter, provided the negative vacuum term has negative pressure.}.

To close this section, we note that perturbations about the
fixed point imply perturbations in $a$ which in turn imply
perturbations in $\rho_\Lambda$. In the case of vacuum energy with $\omega_\Lambda = -1$,
this implies a coupling between matter and vacuum.
Since the conservation of total energy is guaranteed by (\ref{continuity}),
the quantum vacuum would have to exchange energy with matter. We shall consider this coupling in Section \ref{scalarfield} below,
using an alternative approach. We should emphasise here that this
coupling is essential for the above results to hold. This may explain why,
when matter and vacuum are independently conserved and $\omega_\Lambda = -1$,
the stability is achieved only for $\omega_m < -1/3$ \cite{cyclic,EU}.
In the next section we shall explicitly demonstrate this coupling with
the help of a numerical example.

\section{Numerical study of the cosmological dynamics}

\label{numerical}

In this section we make a brief quantitative study of the effects
of the vacuum energy on the dynamics of the universe.
As an example, we consider the case where the energy content consists of
vacuum energy, which we assume to
have $\omega_\Lambda =-1$ and $\rho_\Lambda = C/a^4$,
plus a relativistic matter with an equation-of-state parameter
$\omega_{rm}=1/3$. Using these equation-of-state parameters in Eq. (\ref{secondorder}) we obtain
\begin{equation} \label{scale factor}
3a^3\ddot{a}+3a^2\dot{a}^2+3a^2-2C=0.
\end{equation}

\begin{figure}[t]
\label{figure1}
\begin{center}
\includegraphics[height=3.5cm,width=5.5cm]{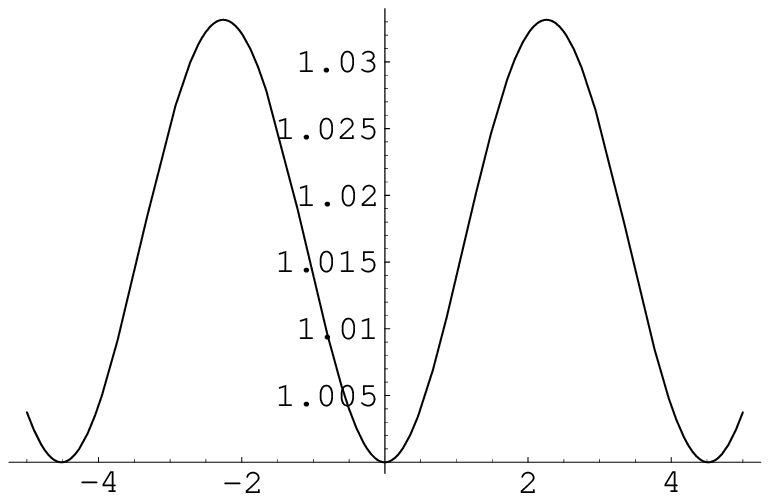}$\;\;\;\;\;\;\;\;\;\;\;\;$
\includegraphics[height=3.5cm,width=5.5cm]{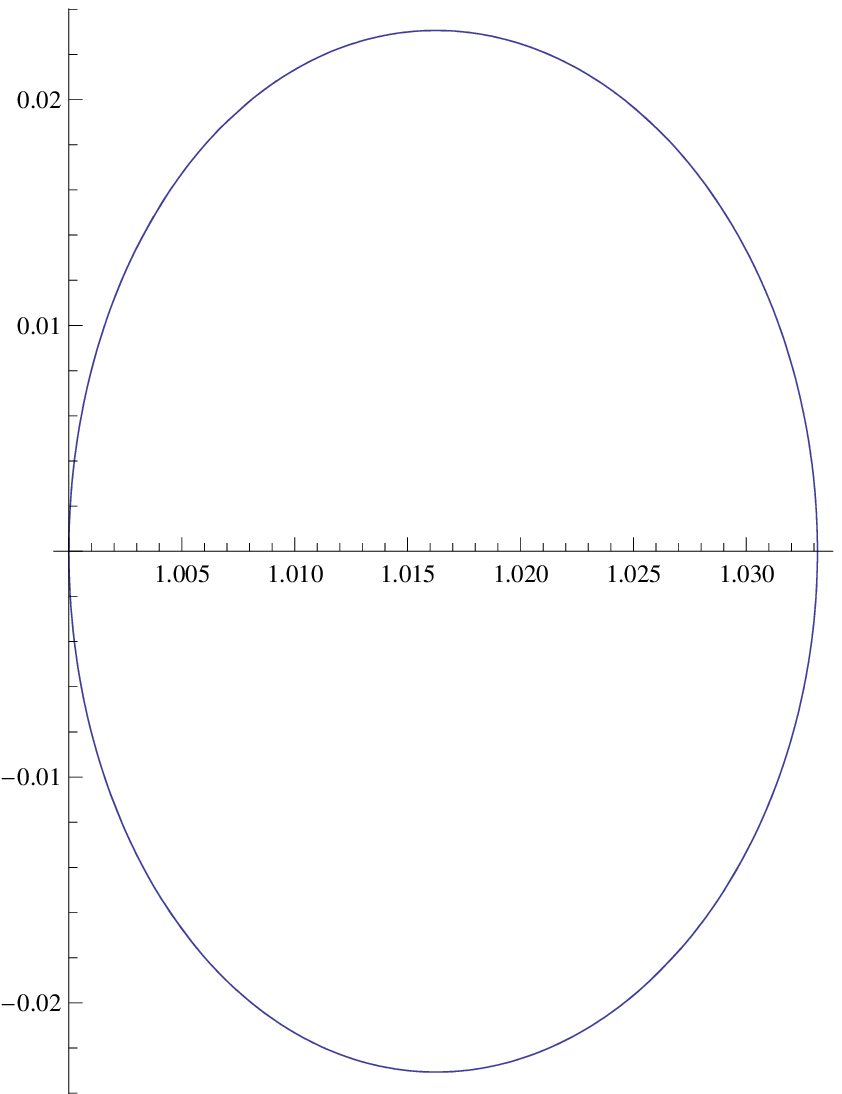}
\end{center}
\caption{The scale factor as a function of time (left) and a typical trajectory
in the phase space ($a, ~\dot{a}$) (right).}
\end{figure}

For the Einstein static solution
the corresponding scale factor is given by
$a^2 = 2C/3$.
Clearly phase space trajectories starting exactly on the
Einstein static fixed point would remain there indefinitely.
On the other hand, trajectories starting in the
neighbourhood of this point would oscillate
indefinitely about this solution.
In Fig. 1 we have plotted an example of
such a universe trajectory using initial conditions
given by $a(0)=1$ and $\dot{a}(0)=0$, together with $2C = 3.1$.

To explicitly demonstrate the coupling between the
matter and vacuum components, discussed above,
we also plot in Fig. 2 the time dependence
of the energy densities of these components.
In the absence of coupling, we would have a constant $\rho_\Lambda$
and a constant $\rho_{rm}a^4$ for relativistic matter.
In our case, however, the effect of the coupling is to make
these quantitities oscillate with time, as can be seen from Fig. 2.

\begin{figure}[t]
\label{figure2}
\begin{center}
\includegraphics[height=3.5cm,width=5.5cm]{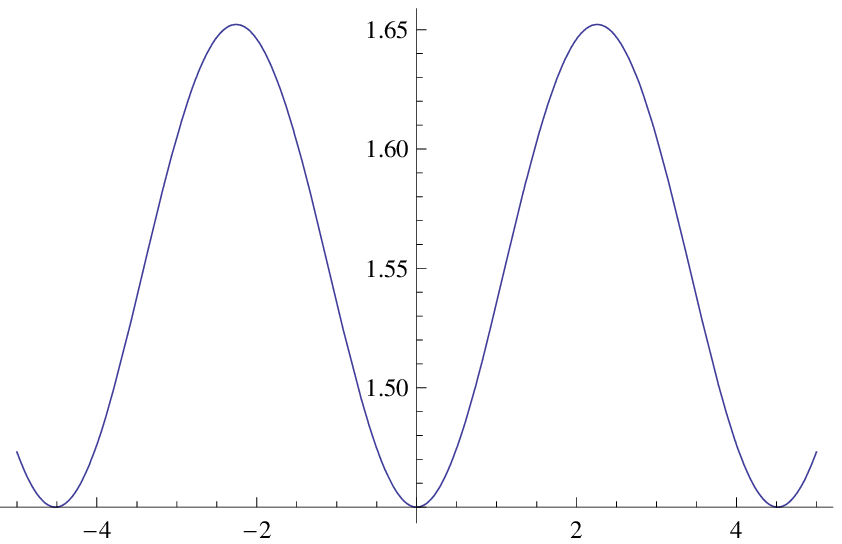}$\;\;\;\;\;\;\;\;\;\;\;\;$
\includegraphics[height=3.5cm,width=5.5cm]{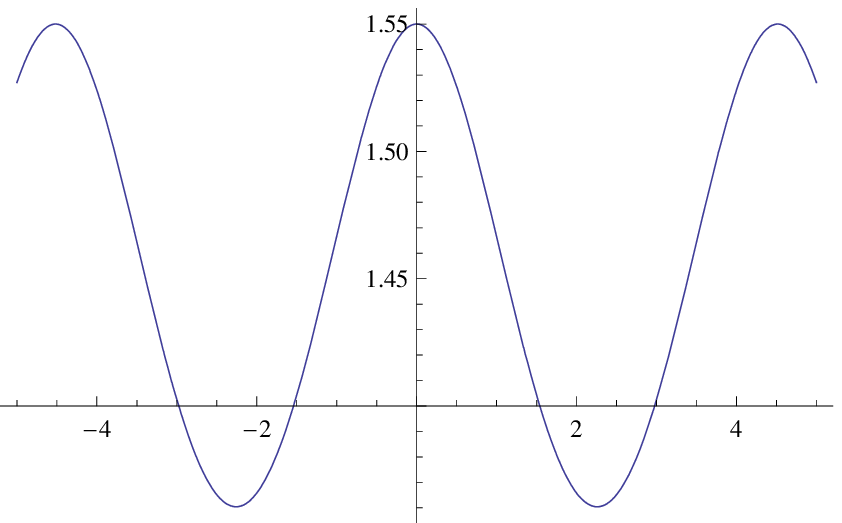}
\end{center}
\caption{The time dependence of $\rho_{rm}a^4$ (left) and $\rho_{\Lambda}$ (right).}
\end{figure}

\section{The scalar field analogy}

\label{scalarfield}

As discussed above, a perturbed Einstein universe with $\rho_\Lambda = C/a^4$ and $\omega_\Lambda = -1$
is only possible if the quantum vacuum exchanges energy with matter.
In this section we consider an alternative procedure,
originally introduced in the case of de Sitter space-time \cite{CT},
in which such a coupling is modelled in terms of a minimally
coupled scalar field $\phi$ with a self-interaction potential $V$.
This can be done by interpreting the usual expressions for the
energy density and the pressure of the scalar field,
\begin{eqnarray}
\label{scalar1}
\rho_\phi &=& \dot{\phi}^2/2 + V,\\ \label{scalar2}
p_\phi &=& \dot{\phi}^2/2 - V,
\end{eqnarray}
as a sum of a vacuum energy component with density $V$ and pressure $-V$,
plus a stiff fluid component with density and pressure equal to $\dot{\phi}^2/2$.
Now, when the scale factor is perturbed, there is an energy exchange between
the potential energy $V$ (here representing the vacuum and taken to be $V =C/a^4$)
and the kinetic energy $\dot{\phi}^2/2$.

The Lagrangian in this case takes the usual form
\begin{equation} \label{Lagrangian}
{\cal L} = \sqrt{-g}\left[\frac{R}{2}-\frac{1}{2}(\partial \phi)^2-V(\phi)\right].
\end{equation}
For a FLRW universe with positive curvature,
the evolution equations take their usual forms
\begin{eqnarray}
3H^2 &=& V + \frac{\dot{\phi}^2}{2} - \frac{3}{a^2}, \label{Friedmann 2}\\
\frac{\dot{\phi}^2}{2}&=&-\dot{H}+\frac{1}{a^2}. \label{scalar field}
\end{eqnarray}
Using (\ref{scalar1})-(\ref{scalar2}),
it is not difficult to cast the above system in the
form of the evolution equations (\ref{Friedmann})-(\ref{continuity}).
Taking for the potential the expression $V =
C/a^4$, we can derive the Raychaudhuri equation in the form
\begin{equation} \label{scale factor 2}
a^3\ddot{a}+2a^2\dot{a}^2+2a^2-C=0.
\end{equation}

The behaviour of this system can be readily studied using the
general analysis given in Section \ref{general}, by recalling that in this
case we have a mixture of a vacuum term (with $\omega_{\Lambda} = -1$) plus
a stiff fluid (with $\omega_m = 1$).
The Einstein static solution again corresponds to $\ddot{a}= 0 = \dot{a}$,
which in this case gives the corresponding scale factor to be
$a^2 = C/2$. A similar analysis to that used above shows that this fixed
point is again a centre.

It is also instructive to briefly re-visit the original classical emergent
model proposed in \cite{EM} in terms of this alternative formulation
of the dynamics in terms of a scalar field.
In that scenario also, the
initial Einstein static phase has a scalar field as its energy content.
We can see from (\ref{Friedmann 2})-(\ref{scalar field}) that, for the
Einstein static solution, $\dot{\phi}^2$ and $V$ are both non-zero constants.
Therefore, while
$\phi$ is changing with time, $V$ is not. In other words, the scalar field
is rolling along a potential plateau. As discussed in \cite{EM}, this
plateau may be considered as the past-asymptotic limit of a smoothly
decreasing potential, which eventually leads to an exit from the Einstein
static regime into an inflationary phase.
Specific forms of such a potential have been considered in \cite{EM,GMT,TL}.

Finally, another possibility that may be considered is that of a complex scalar field.
For example, with a harmonic field $\phi = \phi_0 e^{i\omega t}$ we have
$\dot{\phi}^2=\omega^2\phi_0^2$, and equations (\ref{Friedmann 2})-(\ref{scalar field})
are simultaneously satisfied, with $V = \omega^2\phi_0^2$. Therefore, $V$ remains
constant while $\phi$ rotates in the complex plane. The stability of the solution indicates that this is a local minimum of $V$, and the exit to the inflationary phase may involve, for example, a tunnelling to a global minimum.

\section{Conclusion}

\label{conclusions}

We have studied the existence and stability of
the Einstein static universe in
presence of vacuum energy corresponding
to conformally-invariant fields.
Using the result that the vacuum energy density in
Einstein universe is proportional to the
inverse fourth power of the scale factor,
we have found the range of equation of state parameters
for the vacuum energy such that the Einstein universe
is stable, in the sense of dynamically corresponding to
a centre equilibrium point. The importance of
such a solution is due to the central role
it plays in the construction of non-singular emergent oscillatory
models which are past eternal, and hence can resolve the singularity
problem in the standard cosmological scenario.
\\

Given that the oscillatory universe discussed above
is close to but not exactly an Einstein universe,
the form of the vacuum energy density in the initial
oscillatory phase of the universe may depart
from the above inverse fourth power form.
To partially answer what happens if vacuum density
takes other forms, we considered, as a first step in this direction, a more general functional form of the type
$$\rho_\Lambda = C/a^n.$$
Proceeding in a similar manner to that used above we
have been able to show that,
in the case where the
content of the universe consists
of radiation (with $\omega_m = 1/3$) plus vacuum energy
(with $\omega_{\Lambda}=-1$), the
Einstein static universe still exists and is
stable (is a centre fixed point) if $n>2$.

\section*{Acknowledgements}

Saulo Carneiro was partially supported by CAPES (Brazil).

\end{document}